\documentstyle[epsf,graphicx]{mn}

\def\x2{$\chi^{2}$}

\def\x2{$\chi^{2}$}

\newbox\grsign \setbox\grsign=\hbox{$>$} \newdimen\grdimen \grdimen=\ht\grsign
\newbox\simlessbox \newbox\simgreatbox \newbox\simpropbox
\setbox\simgreatbox=\hbox{\raise.5ex\hbox{$>$}\llap
     {\lower.5ex\hbox{$\sim$}}}\ht1=\grdimen\dp1=0pt
\setbox\simlessbox=\hbox{\raise.5ex\hbox{$<$}\llap
     {\lower.5ex\hbox{$\sim$}}}\ht2=\grdimen\dp2=0pt
\setbox\simpropbox=\hbox{\raise.5ex\hbox{$\propto$}\llap
     {\lower.5ex\hbox{$\sim$}}}\ht2=\grdimen\dp2=0pt

\begin{document}

\title[The angular correlation function of the RASS BSC]
{The angular correlation function of the ROSAT All Sky Survey 
 Bright Source Catalogue}

\author[Akylas et al.]
{\Large A. Akylas$^{1,2}$,  I. Georgantopoulos$^1$, M. Plionis$^1$   \\
$^1$ Institute of Astronomy \& Astrophysics, National Observatory of Athens, 
I Metaxa $\&$ B. Pavlou, Palaia Penteli, 
15236, Athens, Greece \\
$^2$ Physics Department University of Athens, Panepistimiopolis, Zografos,
Athens, Greece \\  
}

\maketitle

\begin{abstract}
We have derived the angular correlation function of a sample of 2096
sources detected in the $\it ROSAT$ All Sky Survey 
Bright Source Catalogue, in order to 
 investigate the clustering properties of AGN in the 
 local Universe. Our sample is constructed by 
rejecting all known stars, as well as extended X-ray sources.
 Areas with $|b|<30^\circ$ and declination $\delta <-30^\circ$ 
 are also rejected due to the high or uncertain 
 neutral hydrogen absorption.
 Cross-correlation of our sample with the Hamburg/RASS 
 optical identification catalogue, 
 suggests that the vast majority of our sources are indeed AGN.
A 4.1$\sigma$ correlation signal between 0$^\circ$ and 8$^\circ$ was detected 
with $w(\theta<8^\circ)=2.5\pm 0.6\times 10^{-2}$. 
Assuming a 2-point correlation function of the form 
$w(\theta)=(\theta/\theta_{\circ})^{-0.8}$, we find 
$\theta_{\circ}=0.062^\circ$.
 Deprojection on 3 dimensions, using the Limber's equation,  
 yields a spatial correlation length of 
$r_{\circ}\approx 6.0 \pm 1.6 \; h^{-1}$ Mpc. This is consistent with 
the AGN clustering results 
derived at higher redshifts in optical  surveys and suggests a
comoving 
model for the clustering evolution.

\end{abstract}

\newpage

\section{Introduction}
Active Galactic Nuclei (AGN) can be detected at high redshift
 due to their high luminosities and hence they can 
 be used to place stringent constraints on the 
 evolution of Large Scale Structure 
 over a wide redshift range (see Hartwick et Schade 1989). 
 Our knowledge on the AGN clustering properties comes  mainly from 
large optical UV excess (UVX) surveys for QSOs (Boyle et al. 1988, 
 Fang \& Mo 1993, Shanks \& Boyle 1994, Croom \& Shanks 1996, 
La Franca,
 Andreani \& Cristiani 1998).
Croom \& Shanks (1996) analysed the clustering properties of the Large Bright  
 Quasar Survey (LBQS) and combined the results with that 
obtained from other QSO surveys including the Durham/AAT UVX sample. They 
derived a clustering length of $r_{\circ}=5.4\pm1.1 \;h^{-1}$ Mpc
at a mean redshift of 1.27. La Franca, Andreani \& Cristiani (1998) 
 derived comparable results ($r_{\circ}=6.2\pm1.6 \;h^{-1}$ Mpc) 
 by investigating a sample of $\sim$ 700 quasars in the 
 redshift range $0.3<z\leq3.2$.  
  Comparison of these clustering results  
in different redshifts rather favours  a  
comoving model for the evolution of clustering.
According to this model the amplitude of the correlation function remains fixed
with redshift in comoving coordinates as the galaxy pair 
expands together with 
the background mass distribution. In contrast, in the stable model 
 for clustering evolution, AGN trace clumps of mass which have 
 gravitationally 
 collapsed  in bound units and have therefore ceased to 
 take part in the general expansion of the universe.  
 However, most of the AGN samples used in the analyses above, 
 come from pencil beam surveys 
 and contain a large fraction of high redshift AGN. 
 Georgantopoulos \& Shanks (1994) studied the correlation 
 function of a sample of about 200 IRAS selected Seyfert galaxies.
 They detect a $2\sigma$ clustering signal on scales less 
 than 10 $h^{-1}$ Mpc. Although their statistics are limited, their 
 results are more consistent with a comoving model.   
 It is evident that there is a pressing need 
 for large AGN samples in the local universe in order to 
 place tight constraints on  
 the clustering evolution in a broad redshift range.  

Analogous clustering studies in X-rays have been  scarce but they are 
potentially interesting as they provide the opportunity to derive information 
on how the X-ray selected AGN trace the underlying mass distribution. The 
first direct study (using redshift information) of the correlation function 
and clustering properties of X-ray selected AGN is that of Boyle \& Mo 
(1993). They studied the local $z<0.2$ AGN clustering properties 
using a sample  from the $\it Einstein$ Extended Medium 
Sensitivity Survey (EMSS). They did not detect clustering at a 
statistically significant level ($\approx1.5\sigma$).
Vikhlinin \& Forman (1995) analysed the angular clustering properties of a set 
of deep $\it ROSAT$ observations. Their angular correlation length 
$\theta_o$  translates, using Limber's equation (Peebles 1980), 
 to a spatial correlation length $r_{\circ}$ which  is 
consistent with that derived in optical suveys. Finally, Carrera et 
al (1998)
using a sample of 235 AGN from two different soft X-ray 
surveys, the $\it ROSAT$ Deep Survey (Georgantopoulos et al., 1996)
and the $\it RIXOS$ survey (Mason et al. 2000) derived the first 
direct  evidence (ie., in 3 dimensions) for clustering 
 in an X-ray selected QSO sample. They derived a low spatial
correlation length between  1.5 and 5.5 $h^{-1}$ Mpc depending 
on the 
adopted model of clustering evolution.

In this paper we present a study of the 
 clustering properties of soft X-ray selected AGN  from the   
$\it ROSAT$ All Sky Survey (RASS). 
 The large number of AGN in this sample provides 
excellent statistics and hence the opportunity to  put stringent constraints
on the clustering properties in the local X-ray Univerce. Moreover,
  comparison with  clustering results at higher redshift in 
 both the X-ray and optical regime  could provide valuable information 
on the evolution of clustering in the universe.

\section{The $\it ROSAT$ sample } 

The data used come from  the $\it ROSAT$ All-Sky Survey Bright Source 
Catalogue, RASSBSC, (Voges et al. 1999).
 The RASSBSC is a subsample of the brightest 18811 sources  
 in the RASS. The RASS covers 98 per 
 cent of the sky  in the  0.1-2.4 keV energy 
range with 30 arcsecs angular resolution. 
 We excluded all sources with
count rate  less than 0.1$\rm cts~s^{-1}$.
 This yields a catalogue which has a {\it uniform} count rate 
 limit over the sky (Voges et al. 1999).  
 We further exclude all sources  
 between the strip (-30,30) degrees in Galactic 
latitude in order to minimize the effect of hydrogen absorption.
Although the majority of the RASS sources are AGN  
there is an appreciable contamination from stars and clusters of galaxies. 
 Therefore in order to study the AGN cross-correlation function 
 we first need  to exclude all stars, groups and clusters of galaxies.  
In principle,  one can utilise 
 the Hamburg Optical Identification Catalogue  
 (Bade et al. 1998) which is an ongoing spectroscopic follow-up 
 program of the RASS sources in the Northern sky,  
 containing objective prism spectra for several thousand sources. 
 However, as the Hamburg catalogue 
 contains optical identifications only for a small subsample of the RASSBSC, 
 we choose to use a different approach to exclude the stellar 
and extended objects from our sample. We have therefore cross correlated 
the RASSBSC with known stellar catalogues, such as the
Smithsonian Astrophysical Observatory Stellar Catalogue 
(http://xena.harvard.edu/software/sao)
and the STScI Guide Star Catalogue (http://www-gsss.stsci.edu).
In addition we have excluded all known $\it ROSAT$ White Dwarfs, Cataclysmic 
Variables, X-ray Binaries and hot (OB) stars. Details of these ROSAT 
 catalogues are given in 
the GSFC HEASARC web page http://heasarc.gsfc.nasa.gov.
The above cross-correlation resulted in the rejection of 6246 stars. 
 The next step is to remove all known clusters and groups of 
 galaxies from our sample. 
 Inspection of the Hamburg  catalogue shows 
that all identified clusters appear to be spatially extended in X-rays.  
 Hence we further exclude all sources 
which have an extension flag value greater than 40
 (see Voges et al. for the definition of the extension flag value). 
Indeed the cross correlation of our sample with the Hamburg RASS catalogue in 
the common ($\approx 8000$ degrees) area shows no remaining clusters.
After the application of the above corrections the vast majority of our 
sample, at least in the areas of the sky covered by the Hamburg catalogue, 
 consists of AGN (86 per cent),  
 while there is still some small stellar and galaxy contamination
(about 10 and 4 per cent respectively). 

\begin{figure}
\mbox{\epsfxsize=9.3cm \epsffile{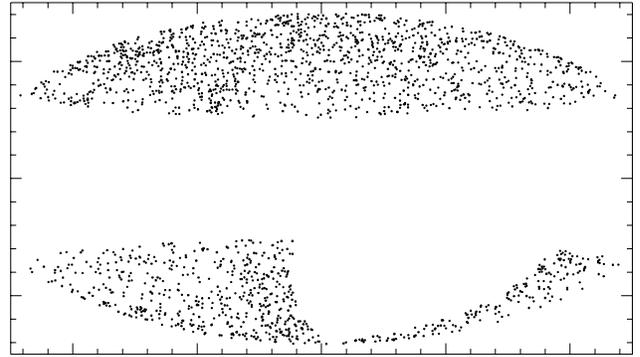}}   
\caption{Aitoff projection of our final catalogue 
 (2096 sources) in Galactic coordinates. The center corresponds to
 Galactic longitude and latitude
l=0, b=0 while l increases to the left}
\label{proj}
\end{figure}

\begin{figure}
\mbox{\epsfxsize=9.3cm \epsffile{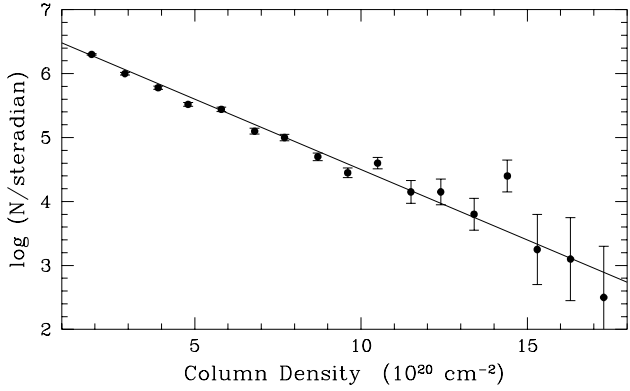}}   
\caption{The AGN surface  density  as a function of hydrogen 
 absorption $N_H$. The errorbars represent Poisson uncertainties.}
\label{nh}
\end{figure}

 Finally, we need to make corrections for the 
   neutral hydrogen in the Galaxy. The 0.1-2.4 keV 
 band is very sensitive to photoelectric absorption.
 Therefore the apparent sky density may change in different areas 
 on the sky due to the changes in the hydrogen column density 
 ($N_H$) in our Galaxy. The large scale variations of the 
 $N_H$ with Galactic latitude as well as the patchy absorption on smaller 
 scales may introduce spurious signals in the correlation function.  
Therefore, we used the Leiden/Dwingeloo Atlas of Galactic Neutral Hydrogen, 
(Hartmann \&  Burton 1997) in order to remove the effects of the hydrogen 
 absorption.  As the above atlas contains $N_H$ data only for  
 declinations $\delta > -30^\circ$ we further exclude 
 RASSBSC sources  outside the missing regions.
 The resulting final catalogue contains 2096 sources over 4.9 sr.  
 In Fig. \ref{proj} we present the Aitoff projection 
 of our sample in Galactic coordinates. 
 The apparent 
large-scale density gradients, seen in Fig \ref{proj}, are due 
to patchy Galactic absorption.
In Fig. \ref{nh} we present the sky density  as a function 
of the Galactic column density $N_H$. 
 The solid line represents the best fit  to 
 the data. This is given by $\log_e N = \alpha + \beta  N_H$ 
 where $N$ is the density of galaxies per steradian and 
 $N_H$ is the column density in units of $10^{20}$ $\rm cm^{-2}$. 
 We find $\alpha=6.7$ and $\beta=-0.2$. 

\section{Method and Analysis} 

We used the two point angular correlation function to measure the clustering  
properties of our sample. 
The correlation function $w(\theta)$ is defined in terms of the 
propability $\delta P(\theta)$ of  finding two objects at a separation $\theta$
in two small solid angles $\delta\Omega_1$ and $\delta\Omega_2$.
                                           
\begin{equation}
\delta P =n^{2}(1+w(\theta))\delta\Omega_1\delta\Omega_2  
\end{equation}
where $n$ is the mean number density of this objects. For an 
unclustered 
population $w(\theta)$=0, while positive or negative values of 
$w(\theta)$ 
indicate clustering or anticlustering respectively.
We estimated the correlation function by comparing the distribution 
of our 
sample to that of a random sample. 
The random sample was constructed by generating 
 angular positions at random over the sky and then 
 folding them through the $N_H$ map.
 In particular, the points on the sky with the lowest $N_H$ are 
assigned a probability 1 (always accepted) while points on the sky with 
 higher $N_H$ are assigned lower probabilities, following 
 the $N$ vs. $N_H$ fit of Fig. \ref{nh}, and therefore have a lower chance 
 of entering the  random catalogue. 
The correlation estimator factor that we use is                               

\begin{equation}
w(\theta)=f(N_{dd}/N_{dr})-1
\end{equation}
 while the error is given by (Peebles 1973) 
\begin{equation}                        
\sigma_{w}=\sqrt{f(1+w(\theta))/N_{dr}}
\end{equation}

Here $N_{dd}$ is the number of data-data pairs 
and  $N_{dr}$ is the number of data-random pairs for a given 
separation.
We use a random sample 30 times larger than our observed catalogue.
 Then the normalization factor is $f=2N_d N_r / N_d(N_d$-1) the ratio 
of the total number of independent data-data pairs to the number of
 data-random pairs. In the equation above $N_d$ and $N_r$ are the 
 number of data and random (after folding through the 
 $N_H$ distribution) points respectively.

\begin{figure}
\mbox{\epsfxsize=8.5cm \epsffile{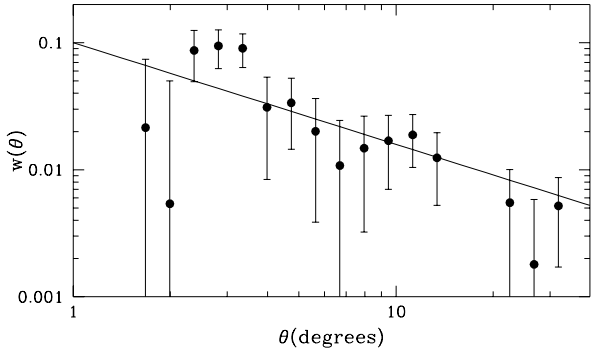}}   
\caption{The two-point angular correlation function $w(\theta)$
 of our sample (2096 sources).}
\label{wthita}
\end{figure}

The derived correlation function is presented in Fig. \ref{wthita}. 
We find $N_{dd}=26669$ and $N_{dr}=26004$ hence detecting 
 a significant clustering at the 4.1$\sigma$ confidence 
level (using Poissonian statistics) between 0 and 8 degrees
 with  $w(\theta<8^{\circ}$)=2.5$\pm 0.6 \times 10^{-2}$. 
On larger scales the correlation function is consistent with an 
unclustered 
population. Assuming a 2-point angular correlation function of the
 form $w(\theta)=(\theta/\theta_{\circ})^{1-\gamma}$ we find that 
$\gamma-1=0.9 \pm 0.15$ and 
$\theta_{\circ}=0.08 \pm 0.03$ degrees. 
If the value of $\gamma-1$ is fixed at 0.8, 
which is the value found for optically selected quasars, we find 
$\theta_{\circ}=(0.06\pm 0.02)$ degrees. 
 Next, we performed a test in order to check whether 
 our signal is diluted due to the small percentage of 
 stellar contamination. We estimated the correlation function for 
all the 
 identified AGN in the 
Hamburg/RASS Catalogue. There were 662 AGN with declination above 
$-30^\circ$. The 
estimated correlation function is $w_{AGN}(\theta)=3.7\pm 1.0\times
10^{-2}$.
 This is consistent with the $w(\theta)$ of our sample, within 
$\sim 1.2 \sigma$, 
 suggesting that the stellar contamination  does not have a major 
 effect on the derived correlation amplitude. 
 We also note that our results do not suffer from 
 the possible 'amplification bias'. Vikhlinin \& Forman 
 (1995) pointed out that when the best-fit angular correlation 
 length is smaller than the FWHM of the ROSAT PSPC Point Spread Function,
  two or more sources can be detected as a single source. 
 As a result  the distribution of the confused sources 
  may be  biased with respect to the 
 correlation function of the parent population. 
 This 'amplification bias' effect  
 works towards the amplification of the 
 correlation function amplitude (cf. Kaiser 1984). 
 However, in our case, confusion problems are very unlikely 
 as the FWHM of the ROSAT PSPC ($\sim 30$ arcsec) 
 corresponds to a very small spatial separation 
 ($<$100 kpc) at the typical redshift of the RASS AGN ($z\sim$0.1-0.2). 

Next, we use Limber's equation which gives the relation between the spatial 
and angular correlation function in order to calculate the correlation  
length $r_{\circ}$ in three dimensions. 
We have used a power-low shape for  the spatial 
two point correlation function  following Groth \& Peebles (1977):
\begin{equation}
\xi(r,z)=(r/r_{\circ})^{-\gamma}~(1+z)^{-p}~,
\end{equation}

\noindent 
where $r_{\circ}$ is the correlation length in comoving 
coordinates, the parameter $\gamma$ is fixed at the
value of 1.8 and $p$ is the parameter describing 
clustering evolution.
If $p=0$ the clustering is constant in comoving 
coordinates (comoving clustering). If $p=1.2$ then the 
clustering is constant in proper coordinates (stable clustering).
The amplitude $\theta_{\circ}$ in two dimensions is related to the 
correlation length $r_{\circ}$ in three dimensions through the equation 
(Peebles 1980):

\begin{equation}
\theta_{\circ}^{\gamma-1}=r_{\circ}^{\gamma}H_{\gamma}
\left(\frac{H_\circ}{c}\right)^\gamma 
\frac{\int_{0}^{\infty} dy \phi(y)^2 [1+z(y)]^{-p}y^{5-\gamma}/F(y)}
{[\int_{0}^{\infty}y^2dy\phi(y)/F(y)]^2}
\end{equation}

\noindent
where $H_{\gamma}=\Gamma(\frac{1}{2})
\Gamma(\frac{\gamma-1}{2})/\Gamma
(\frac{\gamma}{2})$.
The parameter $y$ is related to the  distance and redshift through

\begin{equation}
y=2\frac{(\Omega-2)(1+\Omega z)^{1/2}
+2-\Omega+\Omega z}{\Omega^2(1+z)}
\end{equation}

Finally, $F(y)=[1-y^2(\Omega-1)]^{1/2}$, while 
 $\phi(y)$ determines the fraction of sources observable 
at a given $y$ (or $z$), i.e., 
those with observed fluxes greater than the detection threshold.
The cosmological parameter $\Omega$ enters through the $y-z$ 
relation,
the volume factor $F(y)$ and the selection function $\phi(y)$; here we 
use $\Omega=1$. The selection function 
$\phi(y)$ depends also on the evolution of the AGN luminosity function.
Boyle et al. (1993) derived the AGN cosmological evolution  in the form of 
pure luminosity evolution, with the characteristic luminosity $L_x^\star$ 
varying with redshift as $(1+z)^{3}$. The selection function 
can be written as 

\begin{equation}
\phi(y)=\int_{L_{min}}^{\infty}\Phi(L_x,z)dL
\end{equation}
where the local X-ray luminosity function 
is given by a  double power-law function
(Boyle et al. 1993).

The lower limit of integration $L_{min}$ corresponds to the 
 luminosity that can be observed at redshift $z$ 
 with a flux limit $f_{\rm eff}$. Note that the flux limit 
 changes over the sky although the count rate limit 
 is uniform, ie., $\rm 0.1 \; cts \; s^{-1}$. This is because the 
 flux limit depends sensitively on the column density.
 For the conversion 
 of  the count rate to flux, a spectrum of $\Gamma=2$ was assumed, 
 while the appropriate $N_H$ values 
  were taken from Hartmann \& Burton (1997). 
 The effective RASSBSC flux limit is then 
 the {\it average} flux limit over the areas covered by our sample;
  we find $f_{\rm eff}\sim 10^{-12}$ $ \rm erg \; cm^{-2} \;s^{-1}$.

We derive a value of $r_{\circ}=6.7\pm 1.0 \; h^{-1}$ Mpc 
 assuming comoving clustering evolution and freezing the parameter 
$\gamma$ at the value of 1.8. Alternatively, if we repeat our calculations 
using $\gamma\simeq 1.9$, which we derived from our unconstrained 
$w(\theta)$
fit, we find $r_{\circ}=5.4\pm 0.9 \;h^{-1}$ Mpc. 
We consider the difference 
$\delta r_{\circ} \simeq 1.3 \; h^{-1}$ Mpc between these two determinations 
of $r_{\circ}$ as a further source of 
uncertainty and we quote as our best estimate of $r_{\circ}$ their average 
value: $r_{\circ} \sim 6.0 \pm 1.6 \; h^{-1}$ Mpc. The quoted uncertainty 
is derived by adding in quadrature $\delta r_{\circ}$ and the formal fitting 
errors.
Note that in the case of stable clustering and for $\gamma = 1.8$ we obtain 
$r_{\circ}=6.5\pm 1.0 \;  h^{-1}$ Mpc. 

\section{discussion and conclusions}

Assuming a $\gamma$=1.8 power law model we detect a significant clustering at 
$4.1\sigma$ confidence level between 0 and 8 degrees with 
$w(\theta<8^\circ)=2.5\pm 0.6 \times 10^{-2}$. Our analysis of $w(\theta)$ 
implies 
 $r_{\circ}=6.0\pm 1.6 \; h^{-1}$ Mpc. 
 Note that our result comes from scales larger than 
 $\sim$2 degrees which corresponds to $\sim$ 10 $h^{-1}$ Mpc at 
 a redshift of $z\sim 0.1$. Although the 
 statistics of our sample are limited, the lack 
 of signal at small scales is significant at a high  
 confidence level. This could imply a change in the 
 form of the AGN correlation function at small separations.  
  Our result  represents the first 
significant detection of clustering in the local X-ray universe.
 The Boyle \& Mo (1993) analysis of a sample of 183 low redshift 
($z<0.2$) AGNs, from the EMSS survey, represented the first 
 X-ray selected AGN
clustering study in the local Universe. 
 They  detected a weak but not 
significant clustering signal on scales  $<10^{-1} \; h^{-1}$ Mpc. 

Comparison of our findings with previous results at high redshift can 
 give important clues on the evolution of clustering in the universe. 
 We first compare with the results of Carrera et al. (1998)
 who analysed the clustering properties of two ROSAT, soft X-ray selected 
 AGN samples at higher redshifts, the RIXOS and the 
 DRS samples containing about 235 AGN in total. 
 They detect a 2$\sigma$ clustering signal
  in the RIXOS sample (mean $z\sim 0.8$) while 
 no clustering is detected in the DRS sample. 
 A correlation length of $<$3.5 $\rm h^{-1}$Mpc was found 
 in the case of comoving evolution, significantly lower 
 than our results. However, as their derived 
 correlation length, $r_o$, may depend upon 
    $r_c$ (the distance up to which the number of 
 predicted and observed pairs are compared in order to derive $r_o$),   
 larger samples are needed at 
 high redshifts in order to constrain $r_o$. 
 Better statistics can be provided by using optical samples. 
  Croom \& Shanks (1996) used a sample of $\sim1700$ 
  UVX AGN. They obtain a comoving  clustering length 
 of $r_o=5.4\pm 1.1 \; h^{-1}$ comoving Mpc at a 
 mean redshift of $z=1.3$. Therefore, 
 comparison with our clustering length at low redshifts 
  provides strong evidence for the comoving model of clustering 
evolution.      

Our derived  AGN clustering length is  similar 
 to that of local galaxies. Hence it appears that
 AGN randomly sample the galaxy distribution and they  
 do not trace the high  peaks of the density field 
  despite their low density 
  and high luminosity (cf Efstathiou \& Rees 1988). 
 This result is corroborated by the analysis of 
 the cross-correlation of low redshift EMSS AGN with APM 
 galaxies (Smith, Boyle \& Maddox 1995). 
 They find that the cross-correlation amplitude 
 is similar to the galaxy auto-correlation amplitude 
 thus supporting the idea that AGN randomly sample 
 the galaxy population. This is again consistent 
 with imaging studies of the  AGN environments
 (cf. Boyle \& Couch 1993). These authors find no excess 
 number of galaxies around radio-quiet AGN, 
 again pointing out that galaxies and 
 AGN have similar environments. 

Our work has important implications for the 
 origin of the X-ray background (XRB). 
 Several studies constrained the contribution 
 of AGN to the XRB by using the 
 auto-correlation function of the 
 XRB fluctuations 
 (eg Carrera \& Barcons 1992, Georgantopoulos et al. 1993, 
  Soltan \& Hasinger 1994).
 These studies conclude that if AGN have a clustering  length 
 of $6 \;h^{-1}$ Mpc similar to opticaly selected AGN 
 and they evolve according to the comoving model 
 they can produce only about half of the XRB intensity. 
 Our results appear to maintain the validity of 
 this constraint. 

In the near future, we anticipate a drastic improvement 
 of our knowledge of AGN clustering, at least 
 in X-ray wavelenghts. 
 The RASS identification programmes will 
 be essential in providing a large AGN sample
 in the local universe with redshift information. 
  The new ABRIXAS mission  will be invaluable 
 in providing local AGN samples in the hard X-ray band 
 which is not prone to Galactic absorption. 
 Finally, the XMM serendipitous surveys are expected 
 to provide tens of thousands high redshift AGN 
 thus strengthening the clustering statistics at high redshift. 
 
\section{acknowledgements}
 We are grateful to the referee F. Carrera for many useful comments and 
 suggestions. 
AA thanks Xavier Barcons for many useful suggestions. 
We wish to thank M. Kontiza for her valuable support in the realisation 
 of the above project which is based on a formal collaboration 
 scheme between the University of Athens and the National 
 Observatory of Athens.  
 This research made use of the LEDAS database at 
 the University of Leicester.

\end{document}